\documentclass[10pt,journal,compsoc]{IEEEtran}
\usepackage[normalem]{ulem}
\usepackage[usenames, dvipsnames]{color}
\usepackage{xcolor}
\usepackage{booktabs}
\usepackage{wrapfig}

\usepackage{soul,xcolor}
\setstcolor{purple}

\newcommand{\tm}[1]{}
\newcommand{\cb}[1]{}

\newcommand{\newcopy}[1]{#1}
\newcommand{\remove}[1]{}

\ifCLASSOPTIONcompsoc
  \usepackage[nocompress]{cite}
\else
  \usepackage{cite}
\fi
\ifCLASSINFOpdf
   \usepackage[pdftex]{graphicx}
\else
   \usepackage[dvips]{graphicx}
\fi
\usepackage{url}
\usepackage{hyperref}

\begin{document}
%
\title{Iceberg Sensemaking: A Process Model for Critical Data Analysis}
%
%
%
%

\author{Charles~Berret,
  Tamara~Munzner 
  \IEEEcompsocitemizethanks{
    \IEEEcompsocthanksitem C. Berret is with the division of Media and Information Technologies, Linköping University. Email: charles.berret@gmail.com \protect\\
    \IEEEcompsocthanksitem T. Munzner is with the Department of Computer Science, University of British Columbia. Email: tmm@cs.ubc.ca} \protect\\
   
  \thanks{Manuscript received TK; revised TK.}
}

%
%

\markboth{IEEE TRANSACTIONS ON VISUALIZATION AND COMPUTER GRAPHICS, ~Vol.~xx, No.~x, xxxx~xxxx}%
{Author \MakeLowercase{\textit{et al.}}: Iceberg Sensemaking}
\IEEEtitleabstractindextext{%
  \begin{abstract}
    We offer a new model of the sensemaking process for data analysis and visualization. Whereas past sensemaking models have been grounded in positivist assumptions about the nature of knowledge, we reframe data sensemaking in critical, humanistic terms by approaching it through an interpretivist lens. Our three-phase process model uses the analogy of an iceberg, where data is the visible tip of underlying schemas. In the Add phase, the analyst acquires data, incorporates explicit schemas from the data, and absorbs the tacit schemas of both data and people. In the Check phase, the analyst interprets the data with respect to the current schemas and evaluates whether the schemas match the data. In the Refine phase, the analyst considers the role of power, articulates what was tacit into explicitly stated schemas, updates data, and formulates findings. Our model has four important distinguishing features: Tacit and Explicit Schemas, Schemas First and Always, Data as a Schematic Artifact, and Schematic Multiplicity. We compare the roles of schemas in past sensemaking models and draw conceptual distinctions based on a historical review of schemas in different academic traditions. We validate the descriptive and prescriptive power of our model through four analysis scenarios: noticing uncollected data, learning to wrangle data, downplaying inconvenient data, and measuring with sensors. We conclude by discussing the value of interpretivism, the virtue of epistemic humility, and the pluralism this sensemaking model can foster.
  \end{abstract}

  }

\maketitle

\IEEEdisplaynontitleabstractindextext

%

\IEEEraisesectionheading{\section{Introduction}\label{sec:introduction}}

\IEEEPARstart{D}{ata} analysis is the bedrock of many academic fields, but the responsible use of data means confronting the messy politics of knowledge. Critical studies of data, algorithms, and visualization have grown into a vital interdisciplinary field challenging the authority of data-driven systems \cite{barocas_engaging_2017, boyd_six_2011, boyd_critical_2012, correll_ethical_2019, dork_critical_2013} and calling attention to the many ways these systems fail, \newcopy{skew their subjects, and obscure their inner workings} \cite{oneill_weapons_2016, dignazio_feminist_2016, crawford_atlas_2021}\cb{initial attempt to break up this citation chunk is just to split it in two... 1-5 // 6-8}. Especially troubling cases of algorithmic bias have been tied to search engines \cite{noble_algorithms_2018}, targeted advertising \cite{cadwalladr_revealed_2018}, predictive policing \cite{shapiro_reform_2017, brayne_big_2017}, parole decisions \cite{propublica_machine_2016}, loan approval \cite{eubanks_automating_2018}, and the general reinforcement of existing human prejudice under the guise of mechanical neutrality \cite{benjamin_race_2019, broussard_artificial_2018}. AI-driven systems place data at an even further remove from human analysts, further reducing algorithmic transparency and \newcopy{motivating} calls for critical, humanistic understanding of data-driven reasoning and its limits.

Tracing the source of these critiques leads deep into the philosophical underpinnings of the many fields and industries that increasingly rely on data to make sense of complicated subjects \cite{dignazio_data_2020, jones_how_2021}. With increasing reliance on data comes a \newcopy{tacit} tendency to adopt a \remove{basic, even} theory of knowledge rooted in \textbf{positivism}: the scientific stance that objective truth is accessible through experimental methods based in empirical \newcopy{observation}, and that proper application of these methods will yield results that are independent from the analyst's own perspective and position in the world. Although the basic tenets of positivism are rarely stated aloud, a positivist stance toward knowledge and methodology predominates in visualization research and many of the fields where we \newcopy{most often} collaborate with domain experts \cite{drucker_vis_2019, meyer_criteria_2019}. A clear majority of visualization researchers are at least implicitly positivist in treating large-scale analysis as though it can minimize the subjective dimensions of human judgment. Sometimes this implicit positivism is practical\ and expedient as a default research mode, but it would be an error to imagine that this is how things must be done\remove{, especially in light of mounting evidence that a positivist approach to data can lead us astray}.

Objections \newcopy{and alternatives} to positivism are often grounded in \textbf{interpretivism}, the philosophical stance that knowledge is created through acts of interpretation, that many different interpretations of a given subject are \newcopy{viable}, and the \newcopy{relative} viability of different interpretations can be judged through critical examination of their assumptions, cogency, ethical implications, and explanatory power. For an interpretivist, you cannot separate data from human values, contexts, and practices because doing so is an erasure of essential complexities that shape data and knowledge alike. To resurface these erasures, interpretivist critiques often examine structural power, ideology, and other \remove{matters}\newcopy{social phenomena that may otherwise be} taken for granted as the natural order of things. To be clear, we are not talking about statistical power, but the power enmeshed in structures of society and politics. When aimed at data analysis, the ultimate thrust of interpretivist critique is to see the separation of data from the politics of human knowledge as a fundamental misunderstanding of data itself. 

Our work in this paper is directed at the \textbf{sensemaking} process, a way of modeling the search for new understanding of a subject by gathering data and analyzing it to answer task-specific questions. \newcopy{Although visualization is heavily used in sensemaking, matters of visual encoding and interaction design fall outside the scope of the sensemaking process.} Several influential sensemaking models have been proposed in fields such as \newcopy{HCI \cite{russell_cost_1993}, statistics \cite{grolemund_cognitive_2014}, and visual analytics \cite{sacha_knowledge_2014}}, but these models \remove{evince}\newcopy{follow the typical positivist tendency to minimize how much of human judgment is shaped by social, cultural, and political factors}. Moreover, these models focus on successful sensemaking and largely avoid questions about how sensemaking fails and why. Our core contribution is a new model for sensemaking that is built on interpretivist principles. The Iceberg Sensemaking model starts with the idea that data itself is not the fundamental basis of knowledge, but rather the tip of a larger iceberg constituting frameworks of knowledge, \newcopy{which are} often called `schemas' in the sensemaking literature. As a secondary contribution, we analyze the philosophical baggage inherent in the very concept of a schema, which has been largely unexamined in the sensemaking literature despite serving as a linchpin of past models. 

We propose this new model in the spirit of epistemic pluralism, in which different theories of knowledge should be accorded a basic legitimacy and utility, even those that stand directly at odds. We recognize that positivism has undeniable utility in many research settings, but advocate for the politics of knowledge to be properly checked and examined before defaulting to a positivist approach. This stance fits within the existing pluralist scope of visualization research, where a growing share of interpretivist work is a vital source of new ideas, concepts, and questions~\cite{meyer_criteria_2019}.

We identify and discuss four features of our model that differentiate it from previous sensemaking models.

Following a distinction made by sociologists and historians of science \cite{polanyi_TK_1958, collins_TK_2012, jones_TK_2013}, we argue that the schema, a longstanding feature of sensemaking models, needs to be considered at two different levels: \textbf{Tacit and Explicit Schemas}. From an interpretivist perspective, this distinction points to a difference between the explicit schema provided as documentation or annotation of a dataset, and the tacit schema encompassing disparate, unstated features relating to the creation, context, interpretation, or implications of a dataset. Distinguishing between the tacit and explicit schema allows for both descriptive analysis of and prescriptive guidelines for the interplay between these two parts of a schema: the part we actively work with (explicit) and the part that remains unexamined until it is identified and articulated (tacit).

A second feature of our model is to place \textbf{Schemas First and Always} in sensemaking. Everyone carries tacit schemas into the sensemaking process through past experience, social conditioning, and other factors that guide  interpretation of a given problem. Schemas come into play before even a scrap of data is considered, and these schemas exist throughout the sensemaking process as vital components at every stage. This stance is a key point of distinction with current sensemaking models, where schemas are often treated like they enter the process only after some sensemaking work has already taken place.

The third feature of our model is to treat \textbf{Data as a Schematic Artifact}: every dataset rests on top of a schema that serves as the foundation for the seemingly raw facts and figures it contains. To deploy a familiar analogy, we treat data as the tip of the iceberg in sensemaking. A dataset's explicit schema is visible just \newcopy{above}\remove{below} the waterline, while the tacit schema that guides its original creation and later interpretation is a submerged mass concealed in the deep waters of structural, hegemonic, disciplinary, and interpersonal power \cite{collins_TK_1990}---until and unless we go to the trouble of exploring it. Treating a dataset as objective truth rather than a designed artifact constitutes a failure to map these tacit schemas and thus ignore the schematic structure that has already given shape to \newcopy{the data}. When schemas are explicitly acknowledged and well mapped, many pitfalls of data analysis can be confronted and addressed. Any time an analyst goes to the trouble of scrutinizing the assumptions underlying a particular dataset, they are already treating data as a schematic artifact.

The fourth feature of our model is \textbf{Schematic Multiplicity}, the active consideration of multiple schemas throughout the sensemaking process. Responsible data analysis requires acknowledgment of alternative perspectives. This guiding principle serves as a check on the tendency to carry a single schema all the way to the conclusion without testing it against other possibilities. 

In the following sections, we discuss related work on process models and sensemaking, followed by an in-depth discussion of schemas in different philosophical traditions as a secondary contribution. We then present our main contribution, the Iceberg Sensemaking process model. We then validate the utility and transferability of this model by using it to analyze four data analysis scenarios: noticing uncollected data, teaching data wrangling, downplaying inconvenient data, and measuring with sensors. We conclude by discussing the value of interpretivism and the virtue of epistemic humility---the recognition that knowledge itself is as multitudinous, complicated, and potentially flawed as the human beings who construct that knowledge. We emphasize that epistemic humility does not mean giving up on the pursuit of scientific knowledge, but rather attuning ourselves to its limits and dealing with those limits more frankly. Epistemic pluralism goes hand in hand with epistemic humility. Accounting for the role of interpretation in the sensemaking process opens up consideration for the appropriateness of different theories of knowledge for different use cases in visualization and data analysis. Our deeper point is that positivism is only one approach to sensemaking, and we propose an explicitly interpretivist model as a new alternative, offering inroads to different forms of inquiry and addressing many concerns that positivism cannot.

\section{Related Work}\label{related-work}

A \textbf{process model} classifies actions according to a set of stages, offering practical guidance on the order and transition criteria for progressing through those stages \cite{meyer_nested_2015}. A \textbf{sensemaking model} is a type of process model that centers the human analyst in the act of gathering and interpreting information, capturing the ``how'' of a data analysis workflow, breaking it down into stages, and outlining the transitions that lead to a conclusion. We now \newcopy{summarize}\remove{discuss} three major sensemaking models proposed in previous work. These summaries focus on the role of schemas and related concepts in each model.

\begin{figure*}[!t]
  \centering
  \includegraphics[width=.65\linewidth]{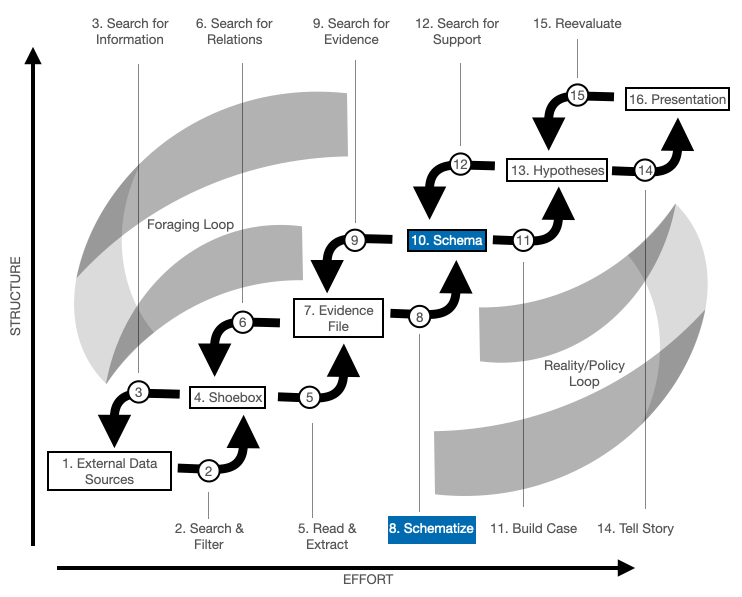}
  \caption{Pirolli and Card (2005) model of sensemaking. The figure has been redesigned to highlight the placement of schemas and schematization.}
\end{figure*}

\begin{figure*}[!t]
  \centering
  \includegraphics[width=.65\linewidth]{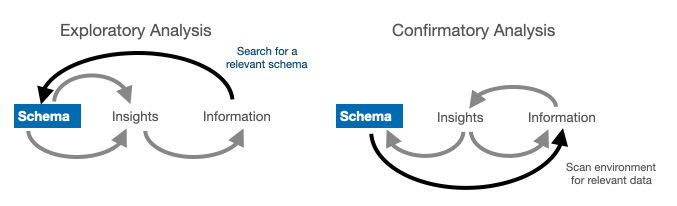}
  \caption{Grolemund and Wickham (2014) model of sensemaking. The figure has been redesigned to highlight the placement of schemas and process of searching for a relevant schema.}
\end{figure*}

\begin{figure*}[!t]
  \centering
  \includegraphics[width=.65\linewidth]{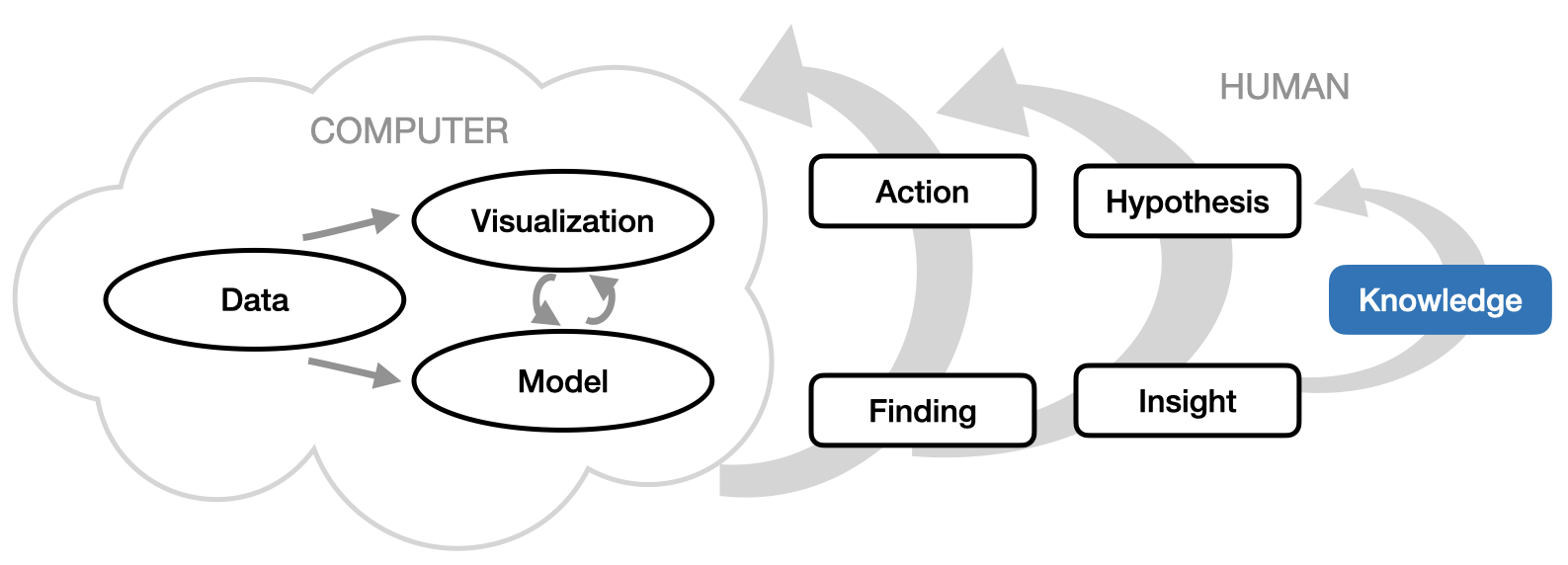}
  \caption{Sacha et al. (2014) model of sensemaking. The figure has been redesigned to highlight the placement of knowledge, the closest approximation of a schema in this model.}
\end{figure*}

\subsection{Sensemaking in HCI and Intelligence Analysis}\label{sensemaking-in-hci-and-intelligence-analysis}

Sensemaking was first described by Xerox PARC researchers Russell, Pirolli, and Card \cite{russell_cost_1993} as the process of ``finding a representation that organizes information to reduce the cost of an operation in an information task'' (p. 271). Pirolli \& Card's \cite{pirolli_sensemaking_2005} subsequent model combines various sensemaking tasks from their earlier work to offer a single, unified articulation of the sensemaking process for intelligence analysis. This model has proved highly influential in a broad set of use cases, particularly in visual analytics. In the Pirolli \& Card model (Fig. 1), the sensemaking process begins by gathering external data sources, then moving elements from these sources into a \emph{shoebox} and \emph{evidence file} before finally formulating a \emph{schema}, leading to a \emph{hypothesis} and ultimately a \emph{presentation} of findings. The process moves forward and cycles back in an iterative refinement sequence, with loops at multiple levels of granularity. Pirolli \& Card treat the construction of schemas as the result of an active, rational, deliberate process in which the analyst fits the information to an apt and useful structure: ``Schemas are the re-representation or organized marshaling of the information so that it can be used more easily to draw conclusions'' (p. 2). As we illustrate below, this treatment of schemas only addresses the explicit schema constructed around a dataset, not the tacit schemas brought to the process by the analyst and implicit in the data itself.

\subsection{Sensemaking in Data Science, Statistics, and Visualization}\label{sensemaking-in-data-science-and-statistics}

Another highly influential model of the sensemaking process covering data science is given by the statisticians Grolemund \& Wickham \cite{grolemund_cognitive_2014}. Their model (Fig. 2) describes the sensemaking process through a parsimonious set of loops in which schemas, insights, and information (data) are the central elements. Grolemund \& Wickham break down the process of data analysis into an exploratory stage and a confirmatory stage. Exploratory data analysis begins with information, then the analyst searches for a relevant schema to explain it. Confirmatory data analysis begins with a schema, followed by a search for data to confirm it. This distinction draws on the statistician Tukey's \cite{tukey_future_1962, tukey_data_1972} seminal models of the data analysis process, which remain central to many modern frameworks for interactive data analysis \cite{hullman_designing_2021}. Grolemund \& Wickham define a schema as ``a mental model that contains a breadth of information about a specific type of object or concept'' (p.5). As we illustrate below, the key distinction between this model and ours is that Grolemund \& Wickham treat schemas merely as mental models and not an active component of the data they work with, postponing the consideration of schemas to the point when the analyst constructs one explicitly.

\subsection{Sensemaking in Visual Analytics}\label{sensemaking-in-visual-analytics}

Sacha et al. \cite{sacha_knowledge_2014} take a visual analytics approach to sensemaking (Fig. 3) that builds upon ideas from Pirolli \& Card to place the interplay between visualization, models, and data as interrelated means of constructing knowledge for the human analyst. Their model also draws upon Norman's \cite{norman_psychology_1988} gulfs of evaluation and execution, defining stages of interactions in which goals lead the analyst from hypotheses to actions (execution) and from findings to insights (evaluation). Their use of the word `knowledge' corresponds to the role of schemas in the other two sensemaking models, and they define it simply as `justified belief'. By treating knowledge (schema) as both the beginning and end point of the sensemaking process, this model is distinct from the previous two because it underlines that human understanding guides every stage of sensemaking. As we later discuss in greater detail, the major limitation of the Sacha model is its under-theorization of knowledge. \newcopy{'Justified belief' is not} an adequate definition, nor is the duality of tacit and explicit knowledge addressed.

\section{Background: Understanding
Schemas}\label{background-three-ways-of-understanding-schemas}

Although the term `schema' is a standard feature of many sensemaking models, it is rarely defined, let alone given close attention as a word with a complicated history and several distinct meanings. We present a brief keyword study \cite{williams_TK_1975} examining how people have thought about schemas and related concepts over time. This broad examination will help delineate different stances toward schemas in philosophy, psychology, and critical theory, calling particular attention to the intersection of these theories with HCI and visualization design. We also briefly describe the use of the word `schema' in the computer science database literature, mainly to distinguish the conventional sense of a database schema from the mental and social schemas that are more fundamental to the sensemaking process. This review spans a substantial historical and disciplinary range, answering a recent provocation from Meyer and Dykes \cite{meyer_criteria_2019} for visualization researchers to give greater attention to the philosophical groundwork underlying the field. 

We group traditional theories of knowledge pertaining to schemas into two broad categories: cognitivist and interpretivist. The principal differences between these perspectives can be summarized as follows. For cognitivists, schemas are discrete, transportable structures of understanding that we acquire in the course of learning and finding order in the world. \newcopy{Cognitivism is the scientific domain from which visualization, as a field, has adopted many of its positivist tendencies}. As we will show in the keyword study below, the cognitivist approach to schemas in the sensemaking literature implicitly adopts a positivist theory of knowledge without acknowledging this paradigmatic baggage. The alternative we offer is interpretivism, which entails viewing schemas as something we construct and acquire through social practices, systems, and institutions that offer a shared means of framing things in the world, albeit one that is highly contingent and contestable. While it is beyond the scope of this article to survey every major touchstone in these disparate theories of knowledge, our focus is to delineate a divergence in the treatment of schemas and articulate an alternative path in visualization and adjacent fields.

In contrast to these two ways of thinking about schemas as mental structures, the word `schema' has a much narrower technical definition in the computer science database literature. A `database schema' is a model describing relationships between entities within a particular dataset or database \cite{garcia-molina_database_2002}. In data analysis, this kind of schema can be explicitly communicated through data dictionaries packaged with datasets, or even through meaningful column headers in tabular spreadsheets. Although this definition is \newcopy{far} from the philosophical concerns outlined below, it carries a historical and procedural connection to these other uses: a database schema is a technical instantiation of a mental schema, much like a schematic drawing in architecture or engineering.


\subsection{Cognitive Schemas}\label{cognitivist-schemas-and-positivism}

The sense of the word `schema' as a mental construct was coined by the philosopher Immanuel Kant, who borrowed it from an ancient Greek word meaning shape or figure. In his \emph{Critique of Pure Reason} \cite{kant_critique_1998}, Kant described schemas as innate forms that \newcopy{determine}\remove{shape} the way perceptible phenomena appear to us. Schemas serve a crucial role in Kant's elaborate theory of knowledge, mediating between our empirical sensation of the world and ideal forms that exist only in our minds. For instance, Kant treats \emph{time} as a schema because it is neither an inherent property of the world nor an abstract concept like a triangle, but rather a mental construct that fundamentally orders our understanding of things. While this origin story may seem \newcopy{historically} distant, it usefully illustrates how the concept of the schema was introduced into modern philosophy and science: schemas \newcopy{act as}\remove{are} ordering principles that easily go unnoticed because they serve as the starting point for knowing anything at all. 

The term `schema' was revived in the early twentieth century by the psychologist Jean Piaget, whose cognitivist theory of mind treats schemas as basic units of understanding that we acquire through inherent cognitive faculties\cite{piaget_language_1955}. For Piaget, schemas enter human cognition at the developmental stage when a child first grasps the permanence of objects. Piaget reasoned that something abstract (a mental schema) must be enabling the child to think about the object in its absence. This theory forms the initial basis of cognitive constructivism in psychology, holding that we acquire schemas over time through basic cognitive faculties and developmental processes. The fundamental power of this theory is its capacity to explain how a child apparently begins as a blank slate and develops into a wielder of various symbols, concepts, models, and categories. This cognitive approach to schemas influenced later models of the human mind derived from cybernetics \cite{wiener_cybernetics_1948, mcculloch_embodiments_1965}, logical positivism \cite{hempel_fundamentals_1952, ayer_logical_1959, carnap_meaning_1958}, formal linguistics \cite{chomsky_three_1956}, and analytic philosophy of mind \cite{fodor_concepts_1998, putnam_mind_1975} to form the interdisciplinary field of cognitive science, where schemas are treated as stable, objectively comprehensible mechanisms of understanding. For a cognitivist, once someone learns the schema for a histogram, truth table, or road map, these are schemas one shares with everyone else who knows them. Nativists take an even stronger view, treating many schemas of perception, reasoning, and even moral judgment as matters of inborn knowledge \cite{mcginn_inborn_2015}.

From the 1950s through the 1980s, psychologists and computer scientists enjoyed a productive exchange of ideas that wedded the cognitive model of schemas to computational problems. In particular, the early AI research of Simon and Newell \cite{simon_sciences_1969, newell_report_1959} worked in pursuit of a general problem solver following the models developed by Turing \cite{turing_computable_1937}, Shannon \cite{shannon_chess-playing_1950}, and others who viewed computers and human minds as essentially the same thing: information processors and pattern recognizers capable of solving discrete problems through symbolic manipulation. Simon treated schemas as computational means of categorizing these problems, first by finding a discrete structure to represent a given problem, then searching for available strategies that would lead to satisfactory solutions. This early AI work was overtly positivist in the assumption that human cognition reduces to the input of objective data, ordered within discrete schemas, and subjected to calculation. Minsky \cite{minsky_TK_1974} as well as Goldstein and Papert \cite{goldstein_TK_1977} recast these schemas as `frames' that order and filter the available information in a given situation.

The application of cognitive science to human-computer interaction (HCI) and visualization led to many notable successes and technical breakthroughs even in these early years \cite{engelbart_TK_1962, kay_TK_1972, papert_TK_1980}. From 1970 forward, some of the most productive interdisciplinary exchanges between psychologists and computer scientists took place at Xerox PARC, where cognitivism informed fundamental R\&D work on computer interfaces \cite{card_TK_1983}. HCI researchers including Shneiderman \cite{shneiderman_future_1982} and Hutchins, Hollan, and Norman \cite{hutchins_direct_1985} recognized the benefits of direct manipulation of computers and often described their findings using a theoretical lens that included cognitive schemas. The foundational work in visualization from PARC closely adhered to a cognitivist framework, as laid out in the Cognitive Coprocessor Architecture \cite{robertson_cognitive_1989}. This approach was followed by the vast majority of early information visualization researchers \cite{card_readings_1999}. Another stream of psychology research involving schemas originates with Gibson \cite{gibson_senses_1966}, who departed from the strictly computational view of cognition with his theory of \emph{affordances}, drawing attention toward the ways a subject may perceive or else fail to perceive different uses of things in their environment. Norman \cite{norman_user_1986, norman_psychology_1988} made fundamental contributions to HCI by adapting Gibson's theory of affordances to design practices, describing the gulf of evaluation and gulf of execution as the key problems to be addressed by designers. \remove{Still, Gibson and Norman present a fairly positivist view of affordances as fundamental properties of objects rather than active constructions of their users.} \newcopy{By viewing affordances as fundamental properties of objects rather than active constructions of their users, Gibson and Norman evince a tacit positivism drawn from the many cognitivist influences on early HCI and Design research.} 

Although visualization as a field has matured and broadened over time, this cognitivist mentality has become the default for development and evaluation of new visual encoding and interaction techniques. \newcopy{Even though many visualization researchers do not identify as positivists, the field has inherited a strong inclination for positivist methods, values, and assumptions through the influence of cognitive psychology.} \remove{With cognitivism has come the historical baggage of positivism, even though many visualization researchers do not identify as positivists.} A recent survey of effective methods in visual data communication \cite{franconeri_science_2021} offers a useful index of progress in this field while also displaying the tendency to view schemas through a purely cognitive lens. Upon presenting a graphic that is just an ambiguous assortment of circles, the authors write: ``If you are having trouble extracting the data from this visualization, it is not your fault---you do not have the needed schema'' (2021, p. 132). This explanation succinctly illustrates the role schemas serve in many cognitive accounts of data visualization today: schemas are treated as discrete, transportable skills that you either have at your disposal, or don't. The uses of a particular schema by different individuals are treated as equivalent, which is undeniably practical in many research scenarios, but comes at the expense of examining the complex social, cultural, and political factors underlying the acquisition, significance, limitations, and variations of these schemas when making sense of data.

\subsection{Interpretive Schemas}\label{interpretivist-schemas-and-frames}

While cognitivism and positivism have yielded valuable research in visualization and HCI, these stances carry unexamined privilege as the assumed basis of the sensemaking process. There is nothing mysterious about the source of this privilege, given the longstanding affinity outlined above, but one consequence of building sensemaking models on these principles has been a general failure to account for context, power, and other entanglements in the construction of knowledge in visualization and HCI.

An \emph{interpretivist} departs from cognitivists and positivists by asserting the variability and contingency of schemas that shape our understanding of things and people in the world. In this domain, schemas may also be called \emph{frames} \cite{goffman_frame_1986}, sharing with Minsky \cite{minsky_TK_1974} as well as Goldstein and Papert \cite{goldstein_TK_1977} an interest in delineating which things fall inside and outside a given frame. Crucial to the interpretivist approach to frames is the recognition that any framework may be \emph{reframed}---challenging its basic assumptions when we find the implications false, unjust, unsound, unfit for the intended purpose, or otherwise lacking. The interpretivist sees empirical data as inseparable from acts of interpretation that occur throughout the process of gathering and analyzing data. Many skeptics of data science readily contend that `raw data is an oxymoron', concealing the interpretive work that already underlies any dataset from the moment we encounter it \cite{bowker_emerging_nodate, gitelman_raw_2013}.

While most early HCI research at Xerox PARC adopted a cognitivist and often positivist lens, as outlined above, the anthropologist Suchman \cite{suchman_plans_1987, suchman_human-machine_2007} departed toward a more interpretivist approach in her research at PARC on the notoriously tricky case of photocopier interfaces. Suchman examined users' interpretive tactics, and especially the breakdown of user understanding, as a result of highly contingent forms of reasoning on the fly while trying to follow instructions for navigating interfaces and other technical tasks. For Suchman, ``crucial processes are interactional and circumstantial, located in the relationships among actors and between actors and their embedding situations'' (2007, p. 30). Here, Suchman departs from the positivist tradition exemplified by Simon and prevalent among other PARC researchers, finding that the schemas guiding understanding and action are highly personal and contextual\cite{suchman_TK_1993}.

Suchman's work has become something of a classic in the interpretivist domain of Science and Technology Studies (STS), an interdisciplinary field in which historians, sociologists, anthropologists, and philosophers examine the construction of facts and artifacts as consequences of human institutions, values, culture, and politics \cite{winner_TK_1980, pinch_social_1984, mackenzie_TK_1999}. Suchman's ethnographic work dovetails with the feminist STS theorist Harraway's \cite{haraway_situated_1988} influential account of \emph{situated knowledge}, which treats any claim to knowledge as a positional standpoint of the individual and their context. The STS literature marks a critical departure from the positivist assumption that knowledge can be treated as objective and human-independent, instead emphasizing interpretive complexities, contextual factors, and systems of domination that shape scientific knowledge and technological systems. 

Another humanistic model of interpretation that dovetails with the STS literature was presented by Drucker \cite{drucker_vis_2019} in a recent IEEE VIS capstone as particularly suitable for critical approaches to visualization: hermeneutics. The basic idea behind \emph{hermeneutics} is that we come to understand the whole of something by examining its parts, and the parts by examining the whole, tracing back and forth in this manner to develop a contextually situated understanding of a subject. This sequential, iterative process of interpretation is often depicted as a \emph{hermeneutic circle}, although its progressive nature is often compared to a spiral instead. Hermeneutics offers a model of understanding centered on individual observers who may construct different interpretations of the same information due to positional differences and other complexities. While many critical theorists have used hermeneutics to emphasize the ultimate instability of meaning through deconstruction, today this stance is largely outweighed by milder forms of hermeneutics focused on the nature and consequences of interpretation as an act that is always partial, personal, and entangled with contextual factors. Drawing on these productive features of hermeneutic analysis, our model depicts sensemaking as an iterative sequence of interpretations tracing back and forth between schemas and data. 

Whereas hermeneutics tends to focus on sensemaking at the individual level, many interpretivists zoom out to the level of \emph{social constructionism}, emphasizing the role of culture, institutions, and power in the creation of meaning \cite{berger_social_1967}. For a social constructionist, even qualities as basic as color, time, and temperature rest on contingent, subjective schemas for collectively making sense of the world around us. To be a social constructionist does not require denying that these qualities ultimately emerge from physical reality, but rather insists that our subjective understanding precedes and makes possible any pursuit of knowledge. Even data that are gathered, ordered, calculated, and otherwise manipulated by a machine are also socially constructed because the schemas that are durably embedded in these machines originate in social processes and institutional structures with specific histories. This point is the crux of many social constructionist \newcopy{accounts} of science and technology, especially in the realm of data. Against the claim that facts can speak for themselves, social constructionists often reply that facts speak for the powerful, whose interests are served by a dominant interpretation expressed in hegemonic appeals to common sense and received wisdom\cite{haraway_situated_1988, dignazio_data_2020}. Critical theorists exemplified by Foucault developed this deeper concern with power by describing the specific ways that modern systems of authority assert control by gathering and manipulating information under the guise of objective knowledge \cite{foucault_archaeology_1972, foucault_discipline_1979}. The power to label, sort, and categorize people and things into schemas becomes calcified in \newcopy{social} systems that reproduce themselves through disciplinary institutions such as schools, workplaces, \newcopy{government,} prisons, and even \newcopy{everyday}\remove{mundane} systems of classification \cite{foucault_order_1971, foucault_birth_1994, bowker_star_TK_2000}. \newcopy{In particular,} critical accounts of prisons and the construction of criminality have been influential in \newcopy{recent} critiques of data science, where algorithms for predictive policing and parole recommendations make use of schemas that reproduce the systemic racism of the criminal justice system \cite{shapiro_reform_2017}. When a social construction like `criminality' is treated as a basic fact about people, even a perfectly sound sensemaking process will go awry by overlooking its own political entanglements. The algorithmic recommendations given to judges on a parole board may be presented as the calculated likelihood of a specific person committing another crime based on their demographics and past offenses, yet these algorithms are notably skewed in terms of race because racialized minorities are often over-policed systematically, whereas white-collar crime is rarely even tabulated \cite{oneill_weapons_2016, karakatsanis_punishment_2019}.

\newcopy{It is important to note that the visualization literature contains an ever-growing thread of work that aligns with interpretivist goals and methods, whether or not this work is labelled with that specific term. For example, the extensively adopted methodology of design studies uses the term `qualitative' rather than `interpretivist' but is explicitly non-positivist, stating that transferability rather than reproducibility is the end goal for all aspects of that work, including schemas constructed by researchers during problem abstraction~\cite{sedlmair_design_2012}. Some recent work does invoke interpretivism more explicitly, including deeper considerations of the ethics of collaboration during design studies~\cite{akbaba_troubling_2023} and particularly larger-scope considerations of epistemological frameworks~\cite{meyer_criteria_2019,akbaba_entanglements_2024}.}  

Beyond philosophical disagreements about the relative soundness of cognitivism and positivism versus interpretivism and social constructionism, data-driven fields as a whole require theoretical foundations capable of identifying and mitigating bias, injustice, and the sort of naive \newcopy{materialism}\remove{realism} that allows these failures to be treated as the output of an objective scientific process. While many interpretivist critiques of injustice in data science focus on algorithms and predictive systems like machine learning (ML) and artificial intelligence (AI) rather than human sensemaking per se \cite{ananny_algorithmic_nodate, noble_algorithms_2018, eubanks_automating_2018}, the schemas deployed early in the sensemaking process are often carried over into automated systems, amounting to a form of tacit knowledge that passes unexamined once these systems are up and running \cite{jones_TK_2013}. A critical, interpretivist, and social constructionist approach to sensemaking directs attention to unjust, inaccurate, and otherwise misleading schemas at the earliest stages of data analysis and visualization. The process outlined below illustrates one way of modeling how to productively work with data while thinking like an interpretivist. 

\section{Model}\label{model}

Here we describe our process of devising this sensemaking model, present an overview of how it works, and discuss its four key features.

\begin{figure*}[!t]
  \centering
  \includegraphics[width=.8\linewidth]{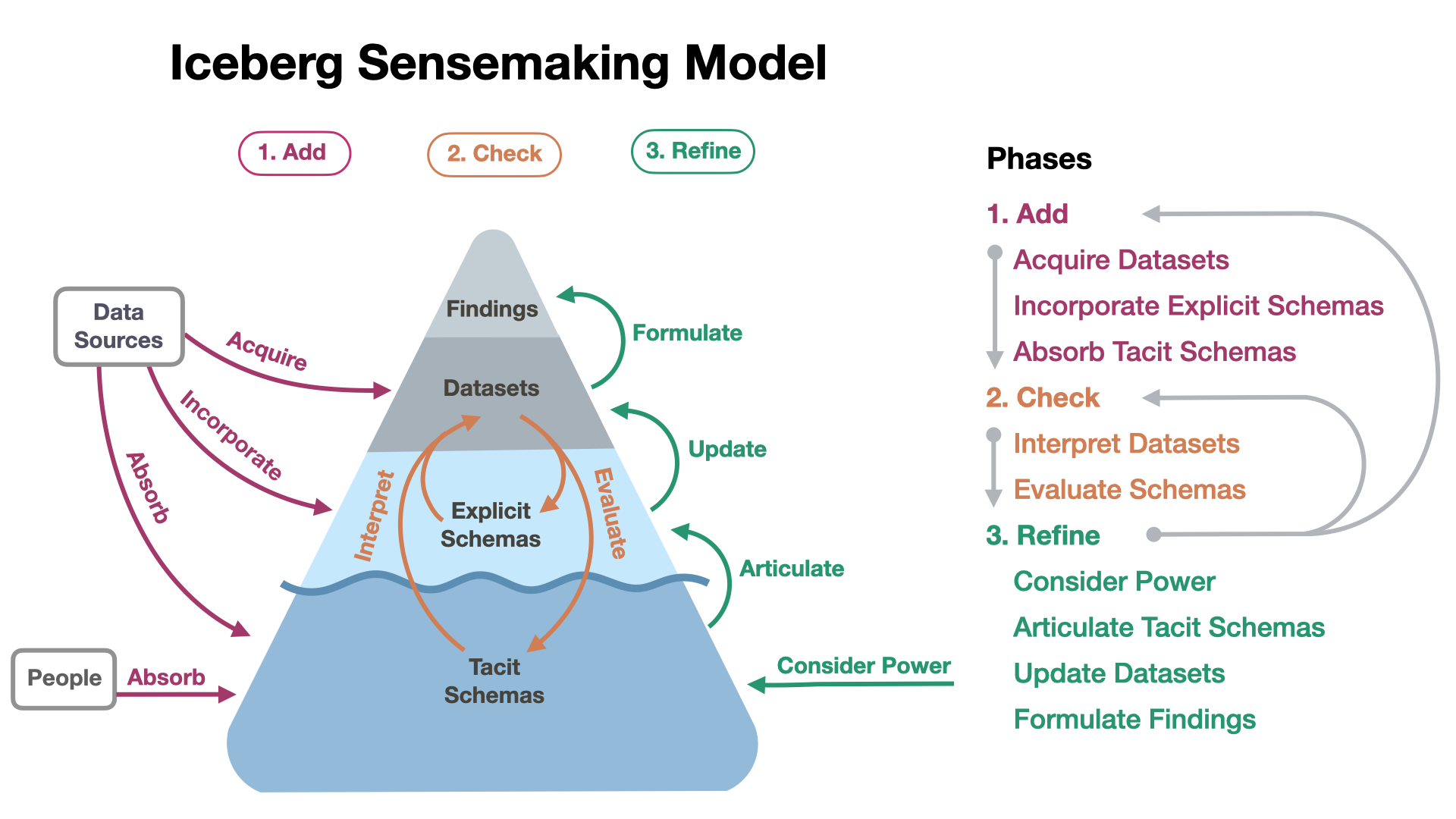}
  \caption{
    \newcopy{The Iceberg Sensemaking Model depicts how a data analyst may use Datasets, Explicit Schemas, and Tacit Schemas to develop Findings through a critical sensemaking process. The wavy waterline separates the Tacit Schemas below from the Explicit Schemas, Datasets, and Findings above. Data Sources and People are external sources of the schemas that eventually become Datasets. The analyst is directed to consider the role of power, in the broad sense of social influence or ideology, to inform their articulation of Tacit Schemas. Actions are classified into three phases (Add, Check, and Refine). The sensemaking process can loop back to any previous phase before reaching a conclusion when Findings are finalized.}}
\end{figure*}

\subsection{Process}\label{process}

This model was constructed through an iterative process of reflective synthesis that included several rounds of literature review. In our initial review of the sensemaking literature, we identified the \emph{schema} as a concept often under-theorized and inconsistently applied, suggesting that a careful examination of schemas could yield insights and new approaches to sensemaking. We then reviewed the literature on schemas, frames, and related concepts to conduct a keyword study of these terms and identify distinctions in their application and implications. Reflecting on these distinctions revealed that our basic assumptions as interpretivists often clashed with the broadly positivist assumptions guiding previous sensemaking models, leading us to develop an alternative sensemaking model from an interpretivist viewpoint with schemas as its core. Early in the model development process, we identified four key principles that constitute fundamental differences in the treatment of schemas within our model versus previous sensemaking models.

We developed and fine-tuned the model itself through multiple rounds of iteration. In each round, we generated sketches of potential structures, then evaluated their descriptive and prescriptive power through detailed walkthroughs of how these models would apply to several sensemaking scenarios. We developed these scenarios to illustrate the practical application of the model to concrete use cases, and to test the clarity and cogency of the developing model. In some cases, we drew directly from the literature, where we found salient critiques of past sensemaking approaches; in others, we devised an emblematic situation to investigate important implications. For each scenario, we describe both a positive and a negative variant. The positive case illustrates functional data analysis, adhering to the principles of the model; the negative case illustrates dysfunctional sensemaking that this model serves to diagnose. When we noted shortcomings in the evolving model's ability to specify important aspects of these scenario contexts or provide guidance to analysts, we honed the model to sharpen its ability to describe and diagnose the sensemaking process in these practical scenarios. In turn, we also refined and augmented the scenarios as the model developed. In Scenarios (Sec. 5), we present these walkthroughs in detail.

\subsection{Overview}\label{overview}

The Iceberg Sensemaking model (Fig. 4) is built around the central metaphor that data is the visible tip of an underlying schematic iceberg. In other words, data always grows out of schemas, not vice versa. The mass of possible schemas always exceeds the dataset given the myriad possibilities of framing data in different ways. We further distinguish between explicit and tacit schemas. The explicit schemas are recorded in some human- or machine-readable way that accompanies what is considered to be the data itself. These explicit schemas can range from annotations, to basic metadata such as column headers, to a complete data dictionary, to a detailed provenance record of how the dataset came to be. Tacit schemas are not explicitly recorded---some are matters of context, some constitute fundamental assumptions in the dataset, and in many cases they are not immediately apparent to the analyst. Tacit schemas include undocumented aspects of the creation and transformation of a dataset, domain knowledge considered common sense by the dataset creators, and unconsidered assumptions that were not surfaced in any previous analysis. If you have ever found the need to stop and examine a previously ignored feature of a dataset that gives it context, you have articulated \newcopy{a}\remove{part of the} tacit schema and \newcopy{made it into an}\remove{moved it into the} explicit schema. Findings are conclusions that have been evaluated, confirmed, and communicated as the \newcopy{final output} of the sensemaking process.

A single iteration of the main sensemaking loop may include any or all of the three phases in sequence: Add, Check, and Refine. Each phase encompasses multiple actions.

In the \textbf{Add} phase, the analyst \emph{acquires} data, \emph{incorporates} explicit schemas from the data, and \emph{absorbs} the tacit schemas of both data and people. At the start of the sensemaking process, an initial dataset is acquired. Acquiring data necessarily results in the incorporation of both explicit schemas and underlying tacit schemas. The absorbed tacit schema of the analyst is their own understanding of the subject matter, which guides them from the beginning of the sensemaking process. The analyst may also absorb the tacit schemas of outside parties, such as managers, colleagues, \newcopy{collaborators,} and other stakeholders. These new schemas may also be specific hypotheses or alternative viewpoints drawn from outside consultation and other information sources, such as news coverage \newcopy{and everyday conversation}.

In the \textbf{Check} phase, the analyst \emph{interprets} the data with respect to their schemas, and \emph{evaluates} whether the schemas match the data. Interpretation means using schemas, both tacit and explicit, to learn about what your dataset contains and draw basic inferences about features in the data, such as trends and outliers. Evaluation means looking at the data to determine whether your schemas are accurate and sufficient to the task at hand. Work conducted in this phase may verify alignment or uncover mismatches between the data and schemas, particularly for newly acquired data or newly incorporated schemas.

In the \textbf{Refine} phase, the analyst should 
\emph{consider} the role of power in their work. This consideration may include questions about who made the dataset, people reflected in the data, potential absences in the data, people affected by the results of their analysis, assumptions made at the outset of the project, or even managerial pressure to produce certain findings. The analyst then \emph{articulates} tacit schemas that were previously unstated. These tacit schemas may arise from any of the three phases: absorbing during the Add phase, evaluating during the Check phase, or 
considering power in the Refine phase. The articulation process moves previously unstated material into active consideration within the explicit schemas. Articulation of tacit schemas is a choice, and the choices an analyst makes about what to make explicit are fundamentally at the core of sensemaking outcomes. Although a complete articulation of tacit schemas is basically impossible because the full context of both the dataset and the analyst's mindset may never be known, the more thoroughly we articulate schemas in conversation with other possible schemas, the greater our certainty that we have not taken for granted false, misleading, or even harmful considerations in the background of the sensemaking process. With this newly articulated material, the analyst may use the explicit schema to \emph{update} the dataset by cleaning, sorting, filtering, recoding, restructuring, and otherwise adapting their datasets. Next, the analyst uses their data to \emph{formulate} findings. These may be preliminary findings that lead the analyst to loop back and perform another Add, Check, or Refine \newcopy{sequence}. Eventually the process ends, perhaps due to deadline pressure or the analyst's subjective sense of satisfaction with the thoroughness of their results. In either case, the final formulation of findings marks the end of the sensemaking process.

Hypothetically, a data analysis scenario could involve only a single pass through the main \emph{Add-Check-Refine} sequence, but like earlier \newcopy{sensemaking} models, the Iceberg Model encourages \newcopy{looping through the process}. Analysts typically iterate with many passes through this loop, repeating the cycle of acquiring and incorporating, interpreting and evaluating, 
considering and articulating and updating, before formulating their final findings. Inner loops may occur at any phase; for example, multiple rounds of interpretation and evaluation could happen before any refinement takes place.

\section{Key Principles}\label{key-principles}

We now expand on the four guiding principles of the Iceberg Model, while highlighting contrasts with previous models.

\subsection{Tacit and Explicit Schemas}\label{tacit-and-explicit-schemas}

Our model reflects the complex dual nature of schemas as both implicit mental models (tacit schemas) and articulated technical frameworks (explicit schemas) that are recorded and accounted for in data analysis. Delineating tacit and explicit schemas allows us to describe and prescribe a process of moving from the former to the latter. Previous models do not capture this aspect of sensemaking.

When sensemaking models use the term \emph{schema} to mean only the explicitly constructed framework of understanding, they fail to guide analysts to articulate and account for tacit background influences on the sensemaking process. When a sensemaking model deploys the term \emph{schema} to mean only the explicit schema, it fails to account for knowledge, beliefs, interests, and biases implicit for both the dataset and the analyst, thus neglecting to address and ameliorate possible sources of bias, power, systemic injustice, and even the basic blindspots of different professions and fields of study.

Pirolli \& Card's \cite{pirolli_sensemaking_2005} treatment of schemas reflects only the creation of the explicit schema, eliding the role of tacit schemas from the beginning of the sensemaking process. The schema plays a larger role in the Grolemund \& Wickham model \cite{grolemund_cognitive_2014}, where it seems to encompass both tacit and explicit dimensions in the search for a relevant schema during the exploratory phase, then the incorporation of new insights during \newcopy{the} confirmation phase. Nevertheless, because Grolemund \& Wickham do not make this distinction, the schema is mainly treated as an explicitly understood model to refine, not an unknown, tacit factor to investigate. The Sacha et al. model \cite{sacha_knowledge_2014} reflects an overly optimistic treatment of the analyst's starting point as a solid block of reliable knowledge. We prefer to treat the assumptions, models, values, and motivations brought to the sensemaking process as schemas in order to underline that these are constructs in need of evaluation and assessment with respect to alternatives.

\subsection{Schemas First and Always}\label{schemas-first-and-always}

Our interpretivist account of sensemaking always begins with schemas. Any account of the sensemaking process that begins without a schema, or some equivalent acknowledgement of the views and perspectives already in play, suggests that the analyst is able to search for data and make sense of a dataset in the absence of inherited values, interests, and frameworks of understanding. Likewise, every dataset carries existing schemas, both tacit and explicit, by virtue of the decisions surrounding the gathering, cleaning, and presentation that preceded the data's arrival on the analyst's desk. Crucially, these initial schemas are just the point of departure: schemas change, develop, and multiply throughout the sensemaking process.

Although the Pirolli \& Card model \cite{pirolli_sensemaking_2005} does discuss schema formation as a framing process, it does not adhere to our \emph{Schemas First and Always} principle. To treat schemas as devices that enter the sensemaking process midway is to foreground the explicit articulation of a schema, ignoring the tacit schemas that shaped the analyst's thinking. This treatment of schemas places a crucial part of the sensemaking process beyond consideration.

Grolemund \& Wickham's \cite{grolemund_cognitive_2014} cognitive model also depicts schemas entering into sensemaking after the process is already underway. Their depiction of confirmatory analysis could be construed as a schema-first process in which one's initially held schema directs the search for relevant data. And yet, in the earlier phase of exploratory analysis, the search for a relevant schema is treated as though it occurs without mental schemas already in play. For a cognitivist who treats schemas as discrete tools that we acquire and place in a kit for later use, it's easy to imagine starting the sensemaking process with some schemas tucked away, others yet to be acquired, and a basically undecided stance in the absence of a working schema. We contend that schemas are the prime mover of the sensemaking process rather than a gear set in motion at a later stage.

Because Sacha et al. \cite{sacha_knowledge_2014} begin and end with \emph{knowledge}, the element closest to a schema in their model, it might seem to meet our \emph{Schemas First and Always} principle. The trouble lies in their treatment of knowledge as a simple input defined as ``justified belief". This definition may seem sensible on its face, but philosophers actually formulate this position as ``justified \emph{true} belief'' and typically use it as the starting point \newcopy{for elaborate arguments against}\remove{to argue against} the absolute need for truth, justification, and even belief as necessary conditions of knowledge\cite{gettier_TK_1963, quine_TK_1969}. Even if we approach justification from another angle, not as an essential property but as a social construction residing in collective agreement, this is not a simple property to locate in one belief versus another, but rather something deeply entangled with structural power from the beginning of the sensemaking process. Crucially, what counts as justification may lead to different sensemaking outcomes, so the criteria for justified belief should be treated as a variegated feature of different schemas introduced throughout the sensemaking process. The \newcopy{key} point is that knowledge is complicated, and it is particularly important for sensemaking models to treat knowledge as complicated.

\subsection{Data as a Schematic Artifact}\label{data-as-a-schematic-artifact}

Data is a designed artifact that arises from choices and acts of interpretation made by those who gather it. We assert that you cannot simply lay schemas on top of data; rather, you need schemas \newcopy{in order} to create data in the first place. Every dataset presupposes a schema that gives it structure and meaning. This principle guides our depiction of data \newcopy{and explicit schemas} as the tip of the tacit schematic iceberg: 
data is the visible outgrowth of a schematic mass that is largely concealed below the waterline. To overlook the primacy of schemas leads to the illusion that data lives a life apart from the human beings that created it.

None of the three major sensemaking models we analyze adhere to this principle. Every previous sensemaking model takes as a starting point the idea that data is a \newcopy{raw} input, fundamentally unschematized in its \newcopy{natural} form. \newcopy{This approach leads schemas to be treated} as the ordering frame that we impose upon neutral facts \newcopy{to make them useful}. The mindset of treating data as \newcopy{a form of material} property, \newcopy{in which} the sensemaking process is intended to make something useful from \newcopy{raw input}, is particularly entrenched in the logic of governments and corporations producing and using datasets. We note that the first sensemaking models emerged from the corporate context of a research lab \cite{russell_cost_1993}, and were later articulated in the governmental context of intelligence analysis \cite{pirolli_sensemaking_2005}.

The mindset of treating data as \newcopy{ground} truth---as raw measurement of objective reality---is very common in the sciences, so a model that sensemaking "uses data to draw conclusions about the world" \cite{sacha_knowledge_2014} feels very natural. The decidedly cognitivist approach of Grolemund \& Wickham \cite{grolemund_cognitive_2014} treats objectivity as the goal, and frames the subjectivity of sensemaking as a flaw to be ameliorated through the data analysis process. In contrast, treating data as a schematic artifact frames data gathering as an intrinsically subjective process\newcopy{, even though it may be useful to treat data as objective in certain methodological contexts}. Our model guides analysts on how to confront the entanglement of data with their own contextual subjectivity, and that of data gatherers, rather than \newcopy{ignore or attempt to eliminate subjectivity}.

Nevertheless, \newcopy{our model does incorporate the utility of the positivist mindset in specific contexts, so we do not wish to abandon it entirely: the sections of the iceberg standing above the water are largely congruent between our model and previous ones}. The difference with the Iceberg model is an expansion of scope that also attends to what happens during the process of gathering the data, before moving on to draw conclusions. Our model insists on the articulation of data's tacit schematization, even if an otherwise positivist stance is responsibly adopted in the sensemaking process. \newcopy{The Iceberg model is a pragmatic effort to synthesize interpretivist approaches within the full scope of existing methodologies, styles of work, and values of visualization as a field.}

\subsection{Schematic Multiplicity}\label{schematic-multiplicity}

A sensemaking model that actively depicts the interaction of multiple schemas better matches the reality of data analysis than a single-schema model. Single-schema models depict the sensemaking process through the interplay of an analyst and their data in a conceptual vacuum, without consulting others or considering alternative perspectives and hypotheses. We specifically call for a sensemaking process that incorporates multiple schemas. At minimum there are two schemas in play, one accompanying a dataset and one in the mind of the analyst. Beyond this bare minimum, we advocate for the explicit consideration of multiple schemas, especially in terms of multiple hypotheses that can be examined through data analysis.

Among the process models discussed above, only the Pirolli \& Card model \cite{pirolli_sensemaking_2005} supports the \emph{Schematic Multiplicity} principle, and it does so only partially. They do allude to the importance of generating multiple hypotheses through annotation of the main model, followed by a brief discussion of this issue, but there is no guidance on how and when to pursue such multiplicity directly within the model itself. The Sacha et al. (2014) model does not explicitly support this principle, nor does Grolemund \& Wickham \cite{grolemund_cognitive_2014}.

While Grolemund \& Wickham's \cite{grolemund_cognitive_2014} approach to data analysis is neatly functional and intuitively powerful, they specifically acknowledge that it suffers from a problem: the potential for an analyst to retain a single false schema throughout the sensemaking process. Our model \newcopy{targets} this tendency with prescriptive guidance to explicitly consider more than one schema.

A single-schema approach to sensemaking lies at the root of many problems in conventional models of data analysis. We argue that these problems can be ameliorated by attending to one's tacitly held schemas up front, explicitly pursuing schematic multiplicity \newcopy{through}\remove{by} considering alternative schemas, and interrogating these schemas throughout the sensemaking process.

\section{Scenarios}\label{scenarios}

We present four scenarios to validate the Iceberg Model's descriptive and prescriptive power \cite{beaudouin-lafon_designing_2004}.

\subsection{Noticing Uncollected Data}\label{noticing-uncollected-data}

The use of data in law enforcement furnishes especially salient examples of the problems that arise when datasets about people and society are framed by the lens of state power.

Consider the case of data-driven policing, in which the messy and biased data produced by law enforcement and the courts often reinforce the policing of race and poverty \cite{oneill_weapons_2016}. A skeptic such as the legal scholar Karakatsanis \cite{karakatsanis_punishment_2019}, who wants to change the public discourse about how we define criminality, makes data-driven arguments to shine light on contradictions and omissions in how the people involved in law enforcement label and prosecute crimes at different levels of society. One of his core arguments is that only some laws, for some people, are enforced.

\subsubsection{Reexamining Lower-Class
Crime}\label{a.-emphasizing-lower-class-crime}

Karakatsanis has illustrated fundamental bias in the US justice system's pursuit, enforcement, and sentencing of crimes typically committed by different socioeconomic classes \cite{karakatsanis_punishment_2019}. Law enforcement data systematically inflates overpoliced, lower-class crimes such as shoplifting, while also deflating numbers for upper-class crimes such as \newcopy{embezzlement and} wage theft that the police are less likely to pursue in their daily work. Using such flawed datasets without attending to their political dimensions amounts to a failure of the data sensemaking process.

Consider a scenario in which a civic official uses police data to make a straightforward assessment of crime trends. In the \textbf{Add} phase, they \emph{acquire} a dataset directly from the local police department showing crime trends over the last six months. The official \emph{incorporates} the explicit schema of this dataset while also \emph{absorbing} tacit schemas from the dataset itself as well as from their own experience.

In the \textbf{Check} phase, the official interprets the data using both the explicit schema that accompanies the dataset, as well as the tacit schemas that frame both the dataset and their own working practices. Likewise, they evaluate these tacit and explicit schemas against the dataset, albeit in a fairly limited manner because they view the dataset as basically reliable. They see nothing especially surprising in these numbers with the same general shape they are used to seeing: high rates of lower-class crimes like shoplifting in poor neighborhoods, and lower overall rates of crime in wealthier neighborhoods.

In the \textbf{Refine} phase, this official does not follow the guidance of the Iceberg Sensemaking model to consider the role of power in the construction of their dataset. As such, they pass over the \emph{articulate} phase without surfacing elements of the tacit schema that might otherwise refine the naive explicit schema they are actively working with. The \emph{update} phase may include some simple data wrangling, but otherwise leaves the original dataset mostly untouched. Finally, the official \emph{formulates} findings based on their dataset, and these findings continue to reflect the general bias of criminal justice data rooted in the over-policing of racialized and lower-class crimes.

\subsubsection{Revealing Upper-Class
Crime}\label{b.-revealing-upper-class-crime}

Now consider a sensemaking process where police data is subjected to critical scrutiny. In the \textbf{Add} phase, an activist \emph{acquires} a dataset providing counts of criminal offenses and incorporates the dataset's explicit schema for crimes such as shoplifting and petty theft as defined by the criminal justice system. They also \emph{incorporate} the tacit schema that gives this dataset its as-yet-unarticulated context as a sociopolitical issue broader than its explicit schema. The activist also \emph{absorbs} tacit schemas from their own experience, as well as schemas drawn from other people such as journalists and researchers whose work informs their general understanding of this subject.

In the \textbf{Check} phase, the activist begins by \emph{interpreting} the data using both their tacit schemas and the explicit schema of the dataset. This initial exploration of the dataset shows that neighborhoods with high rates of poverty are high-crime areas, but in \emph{evaluating} this trend he notes that even though it matches the explicit schema, the phenomenon is more complicated in light of tacit schemas not yet articulated in the data analysis process.

In the \textbf{Refine} phase, they \emph{consider} the role of hegemonic power in the dataset and recognize both omissions and suspect categorizations that skew the dataset toward the initial trends found in the check phase. For example, white-collar crimes such as embezzlement and wage theft are not even tracked, nor are thefts by the police themselves through civil asset forfeiture. By \emph{articulating} these realizations and including them in the explicit schema, the activist either \emph{updates} the dataset to reflect these realizations, or else loops back to the add phase and \emph{acquires} more data about those crimes guided by this new explicit schema. When the activist is satisfied with their dataset, which now includes a more complete accounting of criminal behavior by the rich and powerful as well as the poor, they proceed to \emph{formulate} findings. Their new conclusion is that the financial impact of the white-collar crime dwarfs the measures of theft initially reflected in police data, meaning that rich neighborhoods are the real crime hotspots, not the poor neighborhoods that are over-policed and over-represented in the standard law enforcement datasets. The tacit schema of the original dataset was a mirror reflecting the structure and activity of the criminal justice system, whose over-policing of racial minorities in impoverished locales is mistakenly treated as an objective proxy for the actual totality of criminal activity.

This scenario demonstrates the descriptive and prescriptive power of the Iceberg Sensemaking model. It describes the shortcomings of data used for predictive policing, explains how this system leads to algorithmic injustice through automated systems for predictive policing, and also prescribes improvements through the introduction of alternative schemas reflecting the disproportionate and underrecognized scale of white-collar crime. In this way, sensemaking to uncover data injustice can inform the responsible construction of more just algorithms. It can also be used to call out algorithmic injustice in action. This scenario reflects efforts to rebuke algorithmic injustice by inverting the typical, unjust model of predictive policing and redirecting it toward those who commit crime from a position of privilege \cite{lavigne_TK_2017}. We encourage further critical sensemaking work in other domains in which algorithmic injustice has already been revealed, such as banking \cite{eubanks_automating_2018} and targeted advertising \cite{cadwalladr_revealed_2018}.

\subsection{Learning to Wrangle Data
}\label{learning-to-wrangle-data}

In addition to the broad implications of the Iceberg Model for how we educate data scientists, this model would be immediately effective for recasting lessons on data wrangling \cite{kandel_research_2011, kasica_table_2021} as exercises in scrutinizing the construction and limitations of datasets. \newcopy{The following scenario uses}\remove{This scenario follows} an illustrative \newcopy{classroom} anecdote recounted by boyd \cite{boyd_dragons_2021} in which she encourages data science students to critically explore a dataset before pulling answers from it. Here, the teacher distributes a dataset of police encounters in New York, asking students to find the average age of people apprehended under the city's controversial and highly racialized Stop and Frisk policy. In the course of the lesson, the students first treat the data as ground truth and report a figure skewed by the messiness of the dataset, then update the data after realizing its messy nature. They finally consider aspects of the dataset's construction that might lead to recognizing deeper limitations.

\subsubsection{Getting a Quick Answer}\label{a.-getting-a-quick-answer}

In the \textbf{Add} phase, students first \emph{acquire} the dataset from their teacher and import it into their analysis software. Because the dataset was gathered and published by the New York Police Department (NYPD), the students have \emph{incorporated} the explicit schema of concepts, categories, and demographics used by the NYPD while building this dataset. In addition, they have \emph{incorporated} a range of tacit schemas, including some of the omissions and biases of official data discussed in earlier scenarios. Each student also has absorbed \newcopy{a set of} tacit schemas based on their implicit assumptions and past experience.

In the \textbf{Check} phase, the students begin by \emph{interpreting} the dataset using only its explicit schema, treating it as an objective basis to calculate the average age of people stopped and frisked by the NYPD. The answer yielded by the dataset is 37. The students quickly \emph{evaluate} this figure against their \newcopy{unarticulated} tacit schemas and find the number reasonable.

In the \textbf{Refine} phase, the students initially skip past \emph{consideration} of power and do not \emph{articulate} tacit schemas that may call it into question. They do not \emph{update} the dataset before \emph{formulating} findings based on the raw data, so they reach a straightforward answer to the teacher's question: the students say the average age for offenders recorded in this dataset is 37. The teacher acknowledges that this answer follows directly from the data, but cautions against stopping analysis too soon and encourages the class to loop back and take a closer look.

\subsubsection{Looking Closer at the
Data}\label{b.-looking-closer-at-the-data}

Returning to the \textbf{Add} phase to start a second loop of analysis, the teacher presents a new tacit schema for students to \emph{absorb}, pointing out that extensive media coverage of the stop-and-frisk policy found that teenagers were usually targeted, making 37 a rather high figure for an average. She also presents an explicit schema to \emph{incorporate} into their analysis plan, suggesting that the students should look at the distribution of ages instead of just the mean. The students then jump to the \textbf{Refine} phase, \newcopy{consider power,} and \emph{articulate} this new idea in terms of an explicit schema about the expected age ranges for the targets of this police action.

In a third analysis loop, the students return to the \textbf{Check} phase: the class \emph{re-interprets} the dataset and \emph{re-evaluates} its contents against the new explicit schema. The students notice that many entries record an age of ``99'' and \emph{evaluate} that this number seems implausible, due to a mismatch of the explicit schema on the expected age range and their own implicit schemas on the demographics of the urban citizenry.

Proceeding once more to the \textbf{Refine} phase, the students \emph{articulate} the insight that ``99'' value is likely a placeholder indicating unavailable data, so it should be excluded from calculations, and \emph{update} their dataset to filter them out. They then \emph{re-formulate} findings that reflect a more accurate average age.

Although a basic data wrangling lesson might stop there, the instructor also encourages the students to undertake a fourth analysis cycle, with a self-loop back to \textbf{Refine}, to again \emph{consider} power from a critical perspective in the form of incentives for the people gathering the data. Why were so many ages recorded with the placeholder value? The underlying motivation of police in conducting stop and frisk is to exercise disciplinary power in this domain of surveillance and punishment, so the accuracy of the demographic data may be a lesser concern for them. The police may not choose to devote resources to providing clean data if there is no accountability for ensuring its accuracy, such as an oversight process that ties senior management bonuses to dataset quality metrics. They may even benefit from obfuscating their activities to reduce public scrutiny. The teacher then invites students to investigate the plausibility of other distributions in the dataset, to potentially surface more limitations before formulating their ultimate findings.

Previous sensemaking models would not capture the interplay between data, power, and the various schemas at play in this scenario. Teaching data analysts to thoughtfully engage with politically charged and otherwise fraught subjects such as police data is a challenging remit, but fully tractable beginning with basic lessons like this one and scaling up to more advanced scrutiny of the concepts and categories that fundamentally frame the construction of a dataset.

\subsection{Downplaying Inconvenient Data}\label{downplaying-inconvenient-data}

The Iceberg Model is also useful in focusing attention toward the ways that authority figures have interests that may direct the sensemaking process toward their own preferred conclusions. In some data analysis scenarios, the chain of command may introduce pressure to produce certain data analysis results. Consider the case of an analyst at an investment bank who is asked to evaluate a risk model for the bank's investment strategy. An aggressive strategy that minimizes cash reserves would result in very high yields in the short run for the bank, which would result in a multi-million dollar bonus for their manager. However, it presents substantial risks for the bank's liquidity and stability in the long run. The analyst in the following scenario is under considerable pressure from their manager to conclude that the high-yield path is the best approach.

\subsubsection{Acquiescing to
Pressure}\label{a.-acquiescing-to-pressure}

In the \textbf{Add} phase, the analyst \emph{acquires} datasets based on the bank's standard risk model. They \emph{incorporate} both the explicit schemas for these datasets and the tacit schemas that give these datasets context. They also \emph{absorb} the tacit schema of their manager, who is highly motivated to approve of the strategy that leads to personal enrichment.

In the \textbf{Check} phase, analysis of the dataset shows that the high-risk strategy is unwise. However, the analyst \emph{interprets} the warning as erroneous, because it does not align with their incentives to choose that strategy. They choose not to critically \emph{evaluate} the validity of their schemas, even in the face of this warning.

In the \textbf{Refine} phase, the analyst chooses to change a key assumption in their model, and re-generate a new dataset. In their second analysis loop, returning to the \textbf{Check} phase, the warning disappears. They proceed to the \textbf{Refine} phase and \emph{formulate} findings that the high-risk strategy is sound. The bank profits immensely, and the manager reaps their millions. However, one year later, the bank fails, because the initial warning was in fact correct. This story of motivated reasoning essentially describes what happened with the Silicon Valley bank failure of 2023 \cite{gilbert_silicon_2023}.

\subsubsection{Questioning Authority}\label{b.-questioning-authority}

We now consider how adhering to the Iceberg model could have changed the sensemaking outcome in this scenario. The \textbf{Add} phase proceeds as above. However, in the \textbf{Check} phase, the analyst takes the warning arising from the dataset seriously and \emph{evaluates} whether their schemas harbor faulty assumptions. In the \textbf{Refine} phase, they \emph{consider} the role of interpersonal power in their assumptions, and realize that the pressure from their manager to green-light the pursuit of short-term profits should not lead them to dismiss the warning too hastily. Faced with a mismatch between dataset and explicit schema, they could either interpret the dataset as incorrect or evaluate whether the explicit schema may be incorrect. Informed by their consideration of power, they \emph{articulate} an explicit schema recognizing that very real risks may threaten the long-term viability of the bank itself.

These two scenario variants also illustrate the benefits of schematic multiplicity as a means of ameliorating confirmation bias. If multiple perspectives and alternative framings are brought into consideration, it lowers the likelihood of carrying a false schema from the beginning of the sensemaking process all the way to its conclusion.

A skeptic might say ``this is just the result of bad analysis". In fact, that's the point: other sensemaking models would fail to account for the role of power in the workplace and its effect on sensemaking. \newcopy{Similar intrusions of power on the sensemaking process even occur in the realm of scientific research, where there is considerable pressure to produce positive results.} The absence of political considerations in other sensemaking models means these models are unable to explain cases where power knocks data analysis off the rails.

\subsection{Measuring With Sensors}\label{measuring-with-sensors}

We \newcopy{now}\remove{finally} consider the question of sensemaking with datasets that arise from measurements taken with sensors. At first glance, it may be tempting to assume that \newcopy{mechanical data gathering is}\remove{all such contexts are} prime territory for a positivist framing, where interpretivism is an unnecessary diversion of attention. However, the Iceberg Model has explanatory value in showing how the sensemaking process can fail when bad data is certified as official and reliable while alternative data sources are ignored. Our model also has prescriptive value in showing how to avoid these failures.

\subsubsection{Ignoring Unsafe Conditions}\label{a-ignoring-unsafe-conditions}

The Flint, Michigan water crisis in 2014 rested on a failure to properly test the city's water for the presence of toxins like lead. Official samples were gathered from limited locations in affluent areas, even as utterly miasmic, poisonous brown water poured from faucets throughout the city's poorer areas \cite{goodnough_TK_2016}. Even when a dataset is gathered by law to monitor health conditions, flaws in the data can be routinely overlooked. The city of Flint had been regularly testing water, like every municipality in the United States, but flaws in the data were overlooked and dangerous conditions persisted for years because these flaws were not recognized.

To understand how this failure of sensemaking could have occurred, we analyze the original state of Flint's water monitoring through the lens of the Iceberg Model. In the \textbf{Add} phase, the city \emph{acquired} data about water quality that \emph{incorporated} an explicit schema about where and how to conduct the tests. In the \textbf{Check} phase, the measurements obtained were \emph{interpreted} as indicating acceptable quality levels. In the \textbf{Refine} phase, the city \emph{formulated} findings about water safety. Notably, many of the actions suggested by the Iceberg Model were not carried out.

\subsubsection{Recognizing a Water Crisis}\label{b-recognizing-a-water-crisis}

Imagine a new analyst who takes over water quality measurement in a city like Flint, shortly after governmental officials admit to problems in the water testing regime. In the \textbf{Add} phase, the analyst \emph{acquires} data by taking samples of the city's drinking water using the procedure established by their predecessor. They \emph{incorporate} existing explicit schemas, including directives on where and how often to gather and test the water samples. At the same time, the limitations of this dataset are also \emph{incorporated} as an implicit schema because this is essential context that has yet to be articulated. From our privileged perch as post-hoc observers of this water crisis, we recognize that the full story behind the data has yet to be revealed because it lies implicit in a tacit schema. The tacit schema also includes the analyst's personal knowledge, such as news reports of city residents forced to rely entirely on bottled water because their tap water is undrinkable. This knowledge is \emph{absorbed} into the analysis process as a tacit schema.

In the \textbf{Check} phase, the analyst \emph{interprets} the given data and sees similar figures to those reported by their predecessor. The fact that the dataset shows perfectly safe toxin levels leads them to \emph{evaluate} their tacit schemas and wonder whether reports of unsafe water are exaggerated, or if the dataset itself is flawed. Here, the value of schematic multiplicity is evident in the serious consideration of different hypotheses that may require new data and alternative forms of analysis.

Proceeding to the \textbf{Refine} phase, the analyst \emph{considers} the influence of structural power: they note that the limited water samples are taken at just a few houses with relatively new plumbing in a middle-class area, whereas their city also has many impoverished areas linked to a different segment of the water system. They \emph{articulate} an explicit schema stating that a larger set of samples better representing the city as a whole may yield more accurate figures for water quality. They use this new explicit schema to \emph{update} their dataset to reflect its current omissions. Now they loop back to the \textbf{Add} phase, \emph{acquire} additional data from a more representative sample of neighborhoods, and proceed through the sensemaking process again to interpret that data and evaluate their schemas. In the final \textbf{Refine} phase, they \emph{formulate} findings that reflect previously unreported hazards in the city's drinking water.

Beyond this specific scenario, it is worth noting that political complications frequently emerge in the broader sphere of public health data. The Covid-19 pandemic frequently showcased how debates over the limitations of various datasets and the framing of scientific results amounted to sensemaking complications in which multiple schemas were set at odds, often resulting in contentious reexamination of the sensemaking process and how it led to one conclusion versus another~\cite{lee_viral_2021}.

\subsubsection{Quantifying Rainfall}\label{c.-quantifying-rainfall}

Finally, consider a fairly straightforward data analysis scenario: rainfall sensors used by a scientist to assess drought trends in their state. The sensors are placed at convenient locations to facilitate periodic calibration, and the data is gathered remotely, stored in a spreadsheet, and later analyzed to look for short-term changes and long-term trends.

The process begins with the \textbf{Add} phase, when data is \emph{acquired} from sensors. Acquiring data means \emph{incorporating} explicit schemas, such as measurement parameters and actively documented contextual details like geolocation and calibration data. This phase of the process also involves \emph{absorbing} tacit schemas inherent in the dataset and brought by the analyst, such as their background understanding of the landscape, local ecology, and their instruments. The analyst is familiar with the region, makes sound choices about where to place sensors, and uses their knowledge of local topography to distribute sensors for a representative sample.

In the \textbf{Check} phase, the scientist \emph{interprets} sensor data using an explicit schema specifying standard ranges given historical patterns, as well as tacit schemas involving a range of knowledge, skills, and context. For instance, their interpretation may be guided by tacit schemas in the form of hunches \cite{lin_TK_2023} about anomalies in the data. In this case, the scientist notices that several sensors could be miscalibrated and flags them for examination. The analyst also uses the dataset to \emph{evaluate} whether trends or anomalies may challenge their explicit or tacit schemas. For instance, if the data suggest a sharp decline in rainfall (even without faulty sensors), this data may challenge the scientist's everyday, on-the-ground intuition that precipitation had been fairly consistent with past trends.

In the \textbf{Refine} phase, the scientist \emph{considers} the four possible varieties of power from the matrix of domination \cite{collins_TK_1990, dignazio_data_2020} in their data analysis task. For example, the definition of a drought in their state may be a contested political issue worth taking into consideration. If so, they should \emph{articulate} this matter; however, in this case, structural power in the form of local policies and their implications do not raise any red flags for their sensemaking process. Similarly, they conclude that the other three varieties of power and domination are not immediately relevant to their situation. They \emph{update} the data by removing a few incorrect measurements from their dataset because some of the sensors were miscalibrated. Using this updated dataset, they \emph{formulate} findings.

In this scenario, it is fine to work in the basic mode of positivism, treating measurements as a straightforward proxy for the phenomenon under study, while focusing attention and scrutiny mainly on sensor calibration, the soundness of your model, and accurate presentation of statistical findings. In cases like this \newcopy{one}, positivism is expedient and uncomplicated, granting speed and simplicity to the sensemaking process. The same is true in a case like air-traffic control, where a positivistic sense of certainty is generally preferable to critical doubting of the data\newcopy{, even though the controller should also be attuned to potential signs that they may need to doubt their instruments}. Our basic position is that data analysts must perform some due diligence in the examination of power \newcopy{and the reliability of data} before embracing a positivist stance, rather than simply assuming that a straightforward positivist approach to their subject is warranted. The Iceberg Sensemaking model thus accommodates positivism as a matter of epistemic pluralism, recognizing it as productively compatible within a broader interpretivist framework designed to check for socio-political complications \newcopy{and procedural dysfunction} in the sensemaking process.

For this reason, previous sensemaking models could produce adequate descriptive and explanatory accounts of the rainfall measurement scenario because the role of tacit schemas and power is largely uncomplicated. Yet such uncomplicated circumstances are not ideal for testing the value of a sensemaking model, which should also explain how false and misleading interpretations emerge in problematic cases. While a positivist approach to sensemaking may have descriptive power in well understood cases, it has poor explanatory value in situations where sensemaking requires context, skepticism, and critique---qualities that truly characterize the best scientific work, even if these hunches \cite{lin_data_2021} and other insights happen to remain tacit and unarticulated.

The Iceberg Model would diagnose the original shortcoming of the Flint water quality analysis as the failure to articulate tacit schemas or consider the influence of political and socio-economic factors in the sensemaking process. The placement of sensors could be affected by careless municipal policy, or even the misconduct of public officials. The more political a subject, the more likely that gaps and biases will be present in a dataset. In these cases, the typical positivist approach to data gathering and analysis may treat datasets as reliable and objective even though they are skewed in ways difficult to notice because hegemonic ways of thinking will suggest that the data is reliable and everything is fine. The Iceberg Model offers a means of describing both the unproblematic case of rainwater monitoring above, where the role of political influence is minimal, as well as politically problematic cases like water quality testing, in which sensemaking should include close scrutiny of power.

\section{Discussion}\label{discussion}

We discuss the limitations of our approach, the value of interpretivism as our basic epistemic stance, and the general benefits of epistemic humility in data analysis.

\subsection{Limitations}

A fundamental limitation of all interpretivist approaches, including our own, is that centering subjectivity necessarily limits the reproducibility and replicability of results~\cite{meyer_criteria_2019}. Two analysts who use Iceberg Sensemaking may well come to different conclusions. Our model locates this difference in the acknowledgement of tacit schemas, where the specific information one chooses to articulate from the mass of submerged tacit schemas may be entirely contingent, leading to different sensemaking outcomes between different analysts in varying conditions and contexts. When reproducible results are required, a broadly positivist approach is still tenable under this model because the epistemic pluralism of Iceberg Sensemaking is inclusive of positivism, as well---just as long as the limitations of a positivist approach have been sufficiently scrutinized through the broader interpretivist framework of this model. Moreover, positivist approaches will typically be more time-efficient than the slower interpretivist approaches that demand scrutiny of the received wisdom and prevailing expectations of a scientific field.  

\subsection{The Value of an Interpretivist Model}\label{the-value-of-an-interpretivist-model}

In building this model, we have made an active effort to describe how positivist approaches to data analysis work when they work, while also explaining how they go wrong. A sensemaking model should be just as good at diagnosing how the process fails as describing how it succeeds. While the positivism of past sensemaking models is often \newcopy{implicit}\remove{tacit}, it is most evident in the absence of consideration for faulty and misleading data. Correcting this limitation requires more than just the inclusion of a cleaning phase, because the deepest problems with many datasets only emerge through scrutiny of conceptual frameworks that are \remove{often} taken for granted.

An interpretivist stance toward data requires the acknowledgment that data is ultimately rooted in human ideas, practices, values, and institutions. The positivist tendency to isolate data from these matters of human context (tacit schemas) can be expedient but hazardous. The basic function of our interpretivist model of data analysis is to clearly present how existing uses of data are strengthened, not undermined, by starting from interpretivist principles and attending to interpretivist concerns at crucial stages of the sensemaking process, even if a positivist approach is ultimately utilized.

Our model is built on interpretivist principles because the most striking challenges to data science are based in interpretivist critiques. These critiques of data science have already informed works of investigative journalism \cite{angwin_dragnet_2015, propublica_machine_2016}, the emerging industry of algorithm auditing \cite{oneill_weapons_2016}, and official inquiries into the social and political impact of tech platforms \cite{ftc_bigdata_2016}. Some critics are data scientists, mathematicians, and computer scientists dissenting from within \cite{boyd_critical_2012, boyd_six_2011, boyd_dragons_2021, correll_TK_2019a, correll_TK_2019b, oneil_being_2013, broussard_artificial_2018}; others come from fields like sociology and the digital humanities, where explorations of data-driven methods within critical frameworks have already demonstrated the value of this interdisciplinary synthesis \cite{bowker_emerging_nodate, gitelman_raw_2013, drucker_graphesis_2014, dignazio_data_2020}.

As a rule of thumb, the more political a subject, the more fraught the implications of tacit schemas in data analysis, and the greater the need for interpretivist approaches to sensemaking. For example, court records may seem like straightforward data, but they carry a tacit schema reflecting the entire structure of the criminal justice system. Likewise, calibrating a sensor to measure water quality may seem straightforward, but making use of the readings to assess public health concerns involves acts of interpretation that are deeply embedded in social context. While data entry and automated data gathering could be seen as passive tasks that simply require application of the intended cognitive schema, sensemaking \newcopy{with any dataset requires active attention to the revision of explicit schemas through articulation of tacit schemas. Although biased data is typically discussed in the context of ethics, it is just as much an epistemic matter when} unfairness is rooted in failures of investigative rigor.

In the specific case of visualization, the Nested Model of visualization design~\cite{munzner_nested_2009} can be a useful guide for when to take an interpretivist or cognitivist stance. In the Nested Model, the abstraction layer is where the designer must make judgements about the tasks and data of target users, and this is prime territory for an interpretivist approach. Problem-driven design and requirements elicitation are processes that necessarily involve interpretation, judgment, and subjectivity. In contrast, the lower two levels of the Nested Model are more amenable to a cognitivist approach. At the idiom layer, which deals with visual encoding and interaction design choices, success is often gauged using methods such as controlled laboratory experiments where human performance is measured in terms of time and error. At the algorithm layer, where the goal is to develop automatic methods to instantiate particular idioms, success is often judged through computational benchmarks of system time and storage. Oftentimes these unproblematic evaluation criteria can be safely treated in the objective manner typical of positivism. However, one must be alert to the emergence of cultural, political, and ethical factors that call for full adoption of an interpretivist stance. In short, a stance of epistemic pluralism is especially useful---and a variety of sensemaking models are needed---where different visualization design practices call for different theories of knowledge.

When confronted with a highly political subject, such as the tendency for police data to perpetuate systemic racial violence and inequality, cognitivism tends to be a fairly toothless theoretical stance. Cases of social injustice make an especially pointed case for the limitations of cognitivism when it comes to the messy politics of human knowledge. We argue that a purely cognitive approach to schemas is insufficient for a nuanced model of sensemaking, despite the past success of this theoretical stance in domains of visualization and HCI where cognitivism has been an apt, effective, and relatively unproblematic stance, such as scientific visualization and interface design. In cases where \newcopy{a cognitivist or broadly positivist} approach does prove problematic, an interpretivist stance offers tools for critical intervention in the sensemaking process.

In calling for attention to \newcopy{ethical} dimensions of data science, the Iceberg Model is also aligned with D\textquotesingle Ignazio and Klein's \cite{dignazio_data_2020} salutary call for data feminism, an approach to data that confronts power, embraces pluralism of perspectives, and surfaces the context in which data is collected and used. Far from undermining quantitative reasoning, an interpretivist theory of knowledge yields a more humane, ethical, and defensible stance toward sensemaking with data. We argue that a key dimension of the changes called for by D'Ignazio \& Klein will be to account for the centrality, ubiquity, and variety of schemas at play in any sensemaking process.

\subsection{The Virtue of Epistemic Humility}\label{the-virtue-of-epistemic-humility}

Recognizing the social entanglements \newcopy{of data} entails adopting a fundamental stance of humility with respect to \newcopy{data-driven reasoning}. Facts are never neutral, nobody is truly impartial, and matters of knowledge are inevitably shaped by concepts, measurements, models, and tools of our own creation. But this observation does not mean the world is unknowable---far from it. An interpretivist viewpoint calls attention to the fact that knowledge originates in real human labor, ideas, institutions, and collective memory. This framing \newcopy{implies} knowledge must be situated in human context, treating a multiplicity of perspectives and interpretations as elements of a more comprehensive picture than any one viewpoint \cite{haraway_TK_2018}. In order for knowledge to be constructed soundly and communicated clearly, it must always be \newcopy{carefully} framed. And while every framework leaves something outside its edges, there is value in making the framing process explicit because this underlines the purposeful decisions that lead to one conclusion rather than another. Some biases neatly isolate what matters to us, just as other biases lead us astray.

Epistemology is the study of knowledge itself, and epistemic humility acknowledges how easily human beings can fail or be misled in the pursuit of knowledge. Our interpretivist process model promotes the virtue of epistemic humility by foregrounding the messy complexity of knowledge as an \newcopy{ever-present}\remove{omnipresent} factor in sensemaking. This stance is especially needed in data-driven fields, where data's apparent givenness is implied in the etymology of the very word `data', \newcopy{which is} Latin for `what is given'. Opposition to this misnomer led Drucker to suggest the neologism \emph{capta} to emphasize that information is never just given but rather captured, gathered, ordered, and given meaning through people's actions \cite{drucker_humanities_2011}. Whether \newcopy{researchers}\remove{we} adopt the word `capta' or merely reform our assumptions about the nature of data, it is incumbent on all researchers to heed this distinction, particularly in a field so human-centered as visualization is.

While cognitivism has proved both popular and useful at the intersection of psychology and computer science, as a broadly positivist stance it risks drifting into a reductive view of data as objective facts, transportable across contexts, carrying no perspective of their own. The philosopher Nagel \cite{nagel_view_1989} has critiqued this naive realist stance as the fallacy of a \emph{view from nowhere}, arguing that ``the subjectivity of consciousness is an irreducible feature of reality---without which we couldn't do physics or anything else". Haraway \cite{haraway_situated_1988} labels the same problem as the \emph{god trick}, in which the visual presentation of scientific data gives the impression of an all-encompassing understanding like that of an omniscient deity. When data-driven inquiry goes astray, the cause is often this pursuit of an objective view from nowhere without attending to subjective and socially embedded dimensions of data like its context, construction, and ultimate limitations.

Although positivist research methodologies remain sensible approaches to many scientific subjects, we urge visualization researchers to recognize that \newcopy{this stance is just one option. Taking a different stance of epistemic humility, data is} an artifact that ultimately emerges from human interpretation and is often enmeshed in society and politics. Many visualization researchers already take \newcopy{this stance to some extent, valuing contextual and interpretive factors even in studies with the outward appearance of positivist neutrality. To actively promote epistemic humility would mean letting go of that apparent neutrality and instead taking a pluralist stance toward visualization studies.} The ideal outcome \newcopy{will be} to continue developing the powerful affordances of visualization while embracing a supple understanding of data as contingent, constructed, and contestable.

\section{Conclusion}\label{conclusion}

Data analysis offers undeniably powerful tools for visualization research, but the political complexities of human knowledge present ineluctable, though not intractable, problems. For all the utility and appeal of data as the basis for so much scientific inquiry, the prevailing conception of data as \emph{given} should challenge us to give closer attention to the social impact of data-driven reasoning and humanistic alternatives at hand. Data begins with human sensemaking even when its ultimate impact increasingly emerges from automated systems using algorithms, ML, and AI. Matters of race, gender, policing, and economic inequality pose especially difficult challenges, but there is real promise in acknowledging the fundamental role of human interpretation and the complicating effects of power in the sensemaking process. We thus offer an interpretivist model of sensemaking where datasets grow out of the schemas that give them structure, meaning, and context, not vice versa. We distinguish between explicit and tacit schemas, rejecting the treatment of schemas as monolithic structures that previous models place late in the sensemaking process. We also emphasize the importance of examining multiple schemas to avoid the limitations of a single perspective. We showcase the Iceberg Model's descriptive and prescriptive capacity through four scenarios that map the successes and failures of the sensemaking process to the three phases and nine actions of the model. The result is an interpretivist framework that \newcopy{mobilizes} anti-positivist \newcopy{critiques} of data-centric reasoning \newcopy{without rejecting data analysis altogether}.

  \section*{Acknowledgments}

The authors wish to thank Miriah Meyer, Daniel Weiskopf, Ben Shneiderman, Irina Shklovski, Mark Hansen, Nick Diakopoulos, the UBC InfoVis group, the Vis Collective at Linköping University, \newcopy{and the anonymous reviewers of this paper} for their helpful feedback. Steve Kasica contributed to an early iteration of this model that focused on the work of data journalists. Rosalie Yu provided additional design work for Figure 4. This work was supported in part by a grant from the Wallenberg AI and Autonomous Systems Program (WASP) and by NSERC DG RGPIN-2014-06309.

\ifCLASSOPTIONcaptionsoff
  \newpage
\fi



\bibliographystyle{IEEEtran}
\bibliography{IEEEabrv, references}
%

%







\vspace{100 pt}

\begin{IEEEbiography}
[{\includegraphics[width=1in,height=1.25in,clip,keepaspectratio]{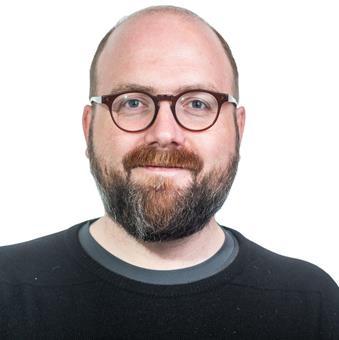}}]
{Charles Berret} received the PhD degree from Columbia University. He is currently a postdoctoral fellow in Critical Data Visualization at Linköping University in Sweden. He was previously a postdoctoral fellow in the UBC InfoVis group. His work has been funded by the Brown Institute for Media Innovation, the Tow Center for Digital Journalism, the Knight Foundation, and the Wallenberg AI, Autonomous Systems, and Software Program (WASP).
\end{IEEEbiography}
\vspace{-33 pt}

\begin{IEEEbiography}
[{\includegraphics[width=1in,height=1.25in,clip,keepaspectratio]{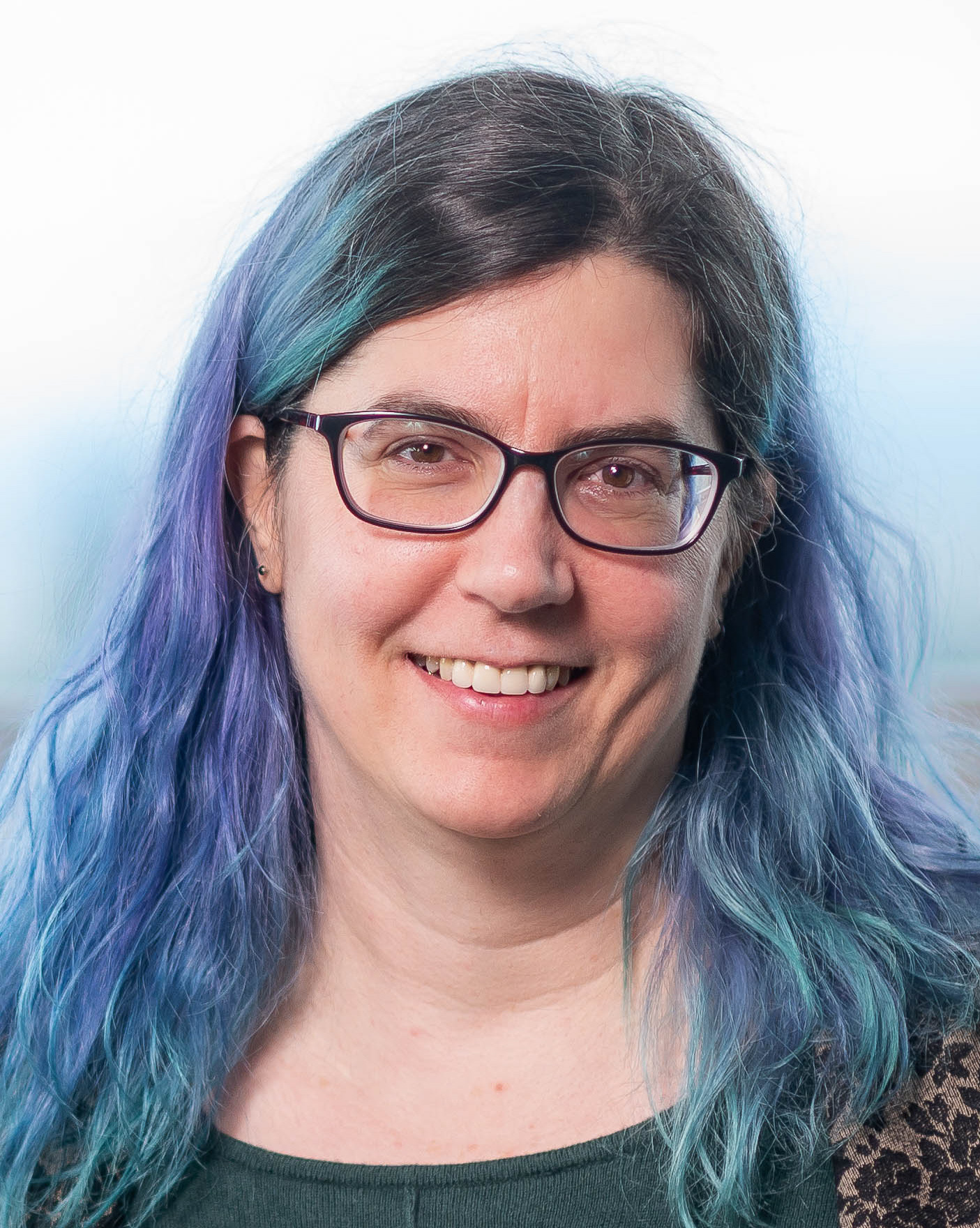}}]
{Tamara Munzner} (IEEE Fellow) received the PhD degree from Stanford. She is a professor with the University of British Columbia. She worked on visualization projects in a broad range of application domains from genomics to journalism. Her book Visualization Analysis and Design is widely used to teach visualization world-wide, and she is the co-editor of the AK Peters Visualization book series at CRC/Routledge. She received the IEEE VGTC Visualization Technical Achievement Award.
\end{IEEEbiography}

\vfill
\end{document}